# Characteristics of natural remanence records in fine-grained particles returned from asteroid Ryugu

Masahiko Sato[1,2,3*], Yuki Kimura[4], Tadahiro Hatakeyama[5], Tomoki Nakamura[6], Satoshi Okuzumi[7], Sei-ichiro Watanabe[8], Seiji Sugita[2], Satoshi Tanaka[3], Shogo Tachibana[2,3], Hisayoshi Yurimoto[9], Takaaki Noguchi[10], Ryuji Okazaki[11], Hikaru Yabuta[12], Hiroshi Naraoka[11], Kanako Sakamoto[3], Toru Yada[3], Masahiro Nishimura[3], Aiko Nakato[13], Akiko Miyazaki[3], Kasumi Yogata[3], Masanao Abe[3], Tatsuaki Okada[3], Tomohiro Usui[3], Makoto Yoshikawa[3], Takanao Saiki[3], Fuyuto Terui[14], Satoru Nakazawa[3], and Yuichi Tsuda[3]

[1]Department of Physics, Tokyo University of Science, Tokyo 162-8601, Japan
[2]Department of Earth and Planetary Science, the University of Tokyo, Tokyo 113-0033, Japan
[3]Institute of Space and Astronautical Science (ISAS), Japan Aerospace Exploration Agency (JAXA), Sagamihara 252-5210, Japan
[4]Institute of Low Temperature Science, Hokkaido University, Sapporo 060-0819, Japan
[5]Institute of Frontier Science and Technology, Okayama University of Science, Okayama 700-0005, Japan
[6]Department of Earth Sciences, Tohoku University, Sendai 980-8578, Japan.
[7]Department of Earth and Planetary Sciences, Institute of Science Tokyo, Tokyo 152-8550, Japan
[8]Department of Earth and Environmental Sciences, Nagoya University, Nagoya 464-8601, Japan
[9]Department of Natural History Sciences, Hokkaido University, Sapporo 001-0021, Japan
[10]Division of Earth and Planetary Sciences, Kyoto University, Kyoto 606-8502, Japan
[11]Department of Earth and Planetary Sciences, Kyushu University, Fukuoka 819-0395, Japan
[12]Graduate School of Advanced Science and Engineering, Hiroshima University, Higashi-Hiroshima 739-8526, Japan
[13]National Institute of Polar Research, Tachikawa 190-8518, Japan
[14]Department of Mechanical Engineering, Kanagawa Institute of Technology, Atsugi 243-0292, Japan

**Key points**
- Stable NRM components were identified in 23 out of 28 Ryugu particles, with paleointensity values ranging from 16.3 to 174 μT.
- Spatially inhomogeneous NRM directions within single particles suggest remanence was acquired before particle solidification.
- NRM characteristics are explained by CRM carried by framboidal magnetite formed during aqueous alteration in Ryugu's parent body.

**Abstract**
Particles collected from the asteroid Ryugu by the Hayabusa2 spacecraft offer a unique opportunity to investigate the magnetic record of the primitive solar system, as any terrestrial magnetic contamination is minimal and can be accounted for. In previous studies, stepwise alternating field demagnetization (AFD) measurements of natural remanent magnetization (NRM) records have been conducted on seven Ryugu particles. However, due to the limited number of samples, there is no consensus regarding the interpretation of the results of these measurements. To address this problem, we performed stepwise AFD measurements of the NRM on twenty-eight Ryugu particles. Twenty-three of the particles exhibited one or two stable NRM components, whereas the remaining five did not. Isothermal remanent magnetization-based paleointensity values derived from stable NRM components varied by more than one order of magnitude. These NRM characteristics were consistent with those observed in previous studies. Therefore, as a reflection of the original nature of NRM record, some Ryugu particles exhibited stable NRM components, whereas others did not. The Ryugu particles investigated in this study and that from a previous study exhibited spatially inhomogeneous NRM directions within individual particles, constraining the NRM acquisition time to before the final solidification of the current Ryugu particles. A mechanism of remanence acquisition that can explain the observed NRM characteristics is a chemical remanent magnetization associated with the growth of framboidal magnetite during aqueous alteration in Ryugu's parent body.

**Plain Language Summary**
This study used tiny rock particles collected from asteroid Ryugu by the Hayabusa2 spacecraft to study the magnetic field records of the early solar system. These particles have not been affected by Earth's environment, making them ideal for this kind of research. This study measured the natural remanent magnetization in 28 particles. Most particles (23 out of 28) held stable magnetic signals, while a few did not. The stable signals varied in strength. Some particles even showed multiple magnetic directions, suggesting the signals were formed before the final solidification of the current Ryugu particles. These magnetic signatures likely came from a process involving water and the growth of tiny magnetic minerals inside Ryugu's parent body. This means the particles preserved a record of the magnetic field that existed very early in the history of the solar system, possibly just a few million years after it began forming.

# 1. Introduction

Knowledge of the dynamic evolution of solar nebula material is key to understanding the history of the solar system. The magnetic field of the solar nebula was generated and sustained as a result of the dynamics of weakly ionized nebular gas, and the materials in the solar nebula dynamically interacted with and coevolved with the magnetic field. Magnetic field strength data from protoplanetary disks obtained from observations of circular polarization produced by Zeeman splitting are limited to regions beyond a few tens of au from the central stars (Vlemmings et al., 2019). For planet-forming regions (approximately <10 au), the only constraints on the magnetic field strength thus far come from measurements of the remanent magnetization of primordial astromaterials of solar system. The natural remanent magnetization (NRM) record obtained from primordial material provides critical constraints on the spatiotemporal evolution of the early solar system (Weiss et al., 2021). To interpret the remanence record and reconstruct the nebular field information precisely, it is essential to characterize the NRM properties of primordial astromaterials.

Ryugu, a C-type near-Earth asteroid (162173), is a rubble-pile body with a spinning-top shape that was formed by the reaccumulation of material ejected from a parent icy planetesimal during catastrophic disruption events (Sugita et al., 2019; Watanabe et al., 2019). The Hayabusa2 spacecraft collected samples from the surface of Ryugu at two locations and transported them to Earth in December 2020 (Morota et al., 2020; Tachibana et al., 2022). Compositional, mineralogical, and isotopic data indicate that Ryugu's parent body and Ivuna-type carbonaceous chondrites (CI chondrites) were formed from the same outer-solar-system reservoirs (Hopp et al., 2022; Kawasaki et al., 2022; Paquet et al., 2022). The Ryugu samples offer a unique opportunity to investigate the dynamic evolution of the material formed in the outer solar system based on the magnetic record because they minimize magnetic field contamination on Earth, which can be traced. The potential contamination sources prior to sample analyses are summarized in Table S1 in Supporting Information.

NRM measurements were conducted on two particles (A0026 and C0002-4-f) during the initial analysis of the asteroid Ryugu samples (Nakamura et al, 2022). These samples were subjected to heating/cooling cycles reaching up to 100°C under ambient laboratory magnetic field conditions, electron microprobe analysis, and X-ray tomography with synchrotron radiation before NRM measurements. As a result of thermal demagnetization (THD) at 110°C and stepwise alternating field demagnetization (AFD), stable components oriented toward the origin in the orthogonal vector plot were recognized in these samples (Nakamura et al., 2022). A preliminary paleointensity estimation based on an isothermal remanent magnetization (IRM) paleointensity method suggested paleointensity values ranging from 41 to 390 μT (Nakamura et al., 2022; Sato et al., 2022).

Maurel et al. (2024) conducted stepwise AFD measurements on samples C0005 and A0154-a that had not been exposed to measurements with the potential to introduce magnetic contamination prior to NRM measurements. These particles did not exhibit stable NRM components. Maurel et al. (2024) subsequently interpreted that the Ryugu particles originally had no stable components due to a weak to null magnetic field during the remanence acquisition event, and that the origin-trending components of A0026 and C0002-4-f were magnetically contaminated. Mansbach et al. (2024) also used Ryugu samples without potentially contaminating the measurements and conducted stepwise AFD measurements on samples A0397, C0085b, and C0006; in contrast to the results in Maurel et al. (2024), they found that three particles exhibited stable NRM components in up to 20–23.5 mT AFD steps. A0397 exhibited a low-coercivity (LC) component upon demagnetization below 10.5 mT, and a middle-coercivity (MC) component upon that between 11 and 23.5 mT. C0085b exhibited an LC component upon demagnetization below 15 mT and an MC component upon that between 15.5 and 23.5 mT. C0006 exhibited one component upon demagnetization below 20 mT. Mansbach et al. (2024) interpreted that these stable NRM components are viscous remanent magnetizations (VRMs) acquired on the spacecraft after sampling and/or from Earth's field prior to the NRM measurements, and that the difference with the results from Maurel et al. (2024) was due to the time spent in a null field.

Stepwise demagnetization measurements of NRM have previously been conducted on seven Ryugu particles, but there is no consensus as to the interpretation of these measurements due to the limited number of analyses performed. Seven particles ranging from millimeters to sub-millimeters in size are considered insufficient to assess the nature of NRM records. Therefore, additional NRM measurements on more Ryugu particles are necessary to enhance our understanding of the NRM records of the Ryugu particles. Therefore, in this study, stepwise AFD measurements of the NRM of twenty-eight Ryugu particles were conducted. The NRM characteristics from previous studies were reevaluated and compared with those of the twenty-eight Ryugu particles measured in this study. Based on the NRM characteristics obtained thus far, including paleointensity values, this paper discusses the nature of the NRM records in the Ryugu particles.

## 2. Methods
### 2.1 Samples

Hayabusa2 performed two touchdown operations on Ryugu on February 21 (TD1) and July 11 (TD2), 2019 (Morota et al., 2020; Tachibana et al., 2022), and surface samples were collected from these two TD sites in chambers A and C, respectively, of the sample catcher (Tachibana et al., 2022). The TD2 landing site was close to an artificial impact crater (Arakawa et al., 2020) and the TD2 samples were expected to contain subsurface materials excavated by the artificial impact. The pebbles and sand inside chambers A and C are representative samples of Ryugu at the two landing sites (Tachibana et al., 2022). The Ryugu particles were passed through and stored on components made from nonmagnetic aluminum alloys (Sawada et al., 2017). The distances from the Ryugu particles to the magnetic and electrical components of the Hayabusa2 spacecraft were larger than 0.4 mm when passing through the sampler horn, 4 mm in the sample catcher transferring into a capsule, and 16 mm after sealing the sample container (Sawada et al., 2017). A temperature monitor attached to the sample container indicated that the container was not heated above 65°C (Yada et al., 2022). The Ryugu particles in the sample container were subjected to a magnetic field of several tens of microtesla from the magnetic moment of the ion engine system (~100 $Am^2$) (Yoshikawa et al., 2016), which was separated from the sample container by ~1.6 m, during transport from Ryugu to Earth. After arrival on Earth, the particles were stored in a curation facility at the Japan Aerospace Exploration Agency.

Magnetic measurements were performed on the particles in chambers A and C. A summary of the sample information and measurement sequences for this study is presented in Table 1. Two sets of samples, the initial analysis and AO samples, underwent magnetic measurements. NRM measurements were conducted on five and twenty-three particles from the initial analysis and AO samples, respectively (Table 1). The initial analysis samples underwent field-emission scanning electron microscopy and X-ray tomography analyses before the NRM measurements, whereas particles C0040-FO065a and C0040-FO065b did not. However, prior to the NRM measurements, the AO samples were not subjected to any other analyses that could modify the remanence record. The grain sizes of the analyzed samples were in a range of several hundred micrometers (Figure S1 in Supporting Information), except for the millimeter-sized particle C0040-FO065a. Particles A0225-03 and A0225-05 originally constituted a single particle, which was split into two daughter particles prior to NRM measurements (Figures S1 and S2 in Supporting Information). The initial analysis and AO samples were kept in fixed orientations for approximately 0.5 and 2 years, respectively, in the curation facility. After these periods, the samples were frequently moved and their orientations changed. The samples were stored in a magnetically shielded case for several days to months until immediately before the NRM measurements. The storage durations of the AO samples in the magnetic shield are summarized in Table 2.

*2.2 Experiments and analysis*

This study employed a remanence measurement technique developed for tiny samples (e.g., Tarduno et al., 2006, 2015; Sato et al., 2015; Kato et al., 2018), as the Ryugu samples analyzed here were smaller than a millimeter. The sample holder for the remanence experiment was identical to that used for the single-crystal silicate measurements reported by Kato et al. (2018). An approximately 1 mm diameter pit was drilled in the center of an alkali-free heat-resistant glass square (Eagle XG, Corning) with dimensions 8 × 8 × 1.1 mm. The drilled glass was used as a sample holder after cleaning in 6 M HCl. The samples were placed in the pit and fixed with $SiO_2$ powder. Remanence measurements were conducted using a superconducting quantum interference device (SQUID) magnetometer (model 755, 2G Enterprises) at the University of Tokyo, Japan. The remanence measurement method followed was that used for single-crystal measurements as described by Sato et al. (2015). The sample holder was fixed on the edge of a polylactic acid rod using double-sided tape. The remanence of the rod was measured before and after the sample measurements, and the average remanence of the rod was subtracted to calculate the sample moment.

Paleomagnetic measurements with stepwise AFD treatments were conducted to assess the NRM of the twenty-eight Ryugu particles used in this study (Tables 1 and 3). An IRM-based paleointensity method (Gattacceca and Rochette, 2004) was subsequently performed on sixteen particles (Tables 1 and 3). The sequence of the paleointensity method was as follows: (1) Stepwise AFD treatment was conducted to measure the NRM. (2) IRM was imparted with a field of 2.5 T using a pulse magnetizer (IM-10-30 Impulse Magnetizer, ASC Scientific), and the resultant IRM was measured with stepwise demagnetization treatment similar to that in the NRM sequence. (3) IRM-based paleointensity value was calculated using the slope in the NRM-IRM diagram ($S_{NRM-IRM}$), which was subsequently used to compute $C_{IRM} \times S_{NRM-IRM}$, where the $C_{IRM}$ is IRM-based paleointensity constant calibrated for the Ryugu samples (3318 µT) (Sato et al., 2022). The slope and its standard error were calculated using ordinary least-squares regression.

Following the IRM-based paleointensity method on sample C0023-FC009, thermoremanent magnetization (TRM) was imparted by heating up to 600°C for 5 min and cooling back to room temperature in a 102.7 µT DC magnetic field ($B_{TRM}$), and the resultant TRM was measured with stepwise AFD treatments same as in the NRM sequence. Heating and cooling procedures were conducted in a vacuum (<5 Pa) using a thermal demagnetizer (TDS-1, Natsuhara-Giken). TRM-based paleointensity values (Shaw et al., 1974) were then calculated using the slope in the NRM-TRM diagram ($S_{NRM-TRM}$), which was subsequently used to compute $B_{TRM} \times S_{NRM-TRM}$. Stepwise AFD treatments on the anhysteretic remanent magnetization (ARM) were conducted on samples C0002-40 and C0023-FC009 (Table 1). To evaluate the effect of the THD treatment on the outcome of the IRM-based paleointensity method, stepwise AFD treatments on the original IRM and the IRM after THD at 110°C of sample C0076-FO007 were performed (Table 1).

NRM and IRM measurements on A0026 and C0002-4-f were conducted during the initial analysis of Ryugu particles (Nakamura et al., 2022). These data were reanalyzed in this study; magnetic measurements were previously reported by Nakamura et al. (2022) and were not conducted in this study. Samples A0026 and C0002-4-f underwent electron microscopy, X-ray tomography, and heating/cooling cycles of up to 100°C under ambient laboratory magnetic field conditions before the NRM measurements. One-step THD treatment at 110°C for 5 min was conducted on the NRM and IRM of samples A0026 and C0002-4-f before stepwise AFD treatment to remove laboratory TRM (Nakamura et al., 2022) (Table S2 in Supporting Information).

## 3. Results

Twenty-three of the twenty-eight Ryugu particles that underwent measurements in this study showed either one or two stable NRM components in stepwise AFD treatments (Table 3; Figures 1 and 2). Linear components continuous from the NRM state comprising more than four points and with a maximum angular deviation (MAD) of less than 30° were identified as stable NRM components. Stable NRM components were identified in the AF ranges of 0–20 and 0–27.5 mT for the initial analysis and AO samples, respectively. Thus, twenty-three out of twenty-eight particles showed stable NRM components with similar AF segments, regardless of whether they underwent prior scanning electron microscopy or X-ray tomography analysis.

Rock-magnetic and microscopic measurements on the Ryugu particles revealed that fine-grained framboidal magnetite, coarse-grained magnetite with hopper, plaquette, and spray shapes, and pyrrhotite grains of various sizes are the main remanence carriers (Nakamura et al., 2022: Sato et al., 2022). Electron holography measurements (Kimura et al., 2023) revealed that the framboidal magnetite grains have a single vortex state, which can obtain and retain high-fidelity magnetic recordings over billions of years (Nagy et al., 2017), while micrometer-sized pyrrhotite grains possess a multidomain magnetic structure, which cannot retain a stable remanence over geological time scales. Moreover, fine-grained sulfide particles distributed in the matrix showed no significant magnetic flux, indicating that these particles are not ferromagnetic phases and do not contribute to the remanent magnetization (Kimura et al., 2023). Rock-magnetic measurements have detected pyrrhotite with coercivity up to approximately 1 T (Sato et al., 2022), suggesting the possible presence of ferromagnetic submicrometer-sized pyrrhotite grains. However, considering the coercivity ranges, these grains are not thought to contribute to the observed stable NRM components. Therefore, the stable NRM components of the AF segments ranging from 0 to 27.5 mT were likely carried by framboidal magnetites. Above critical AF levels of approximately 10–30 mT, the NRM vectors of the Ryugu particles tended to behave unstably at large amplitudes (Figures 1 and 2). Because the Ryugu particles contain large pyrrhotite crystals with grain sizes of up to several tens of micrometers (Nakamura et al., 2022), combined with the fact that coercivity ranges of unstable components above 10–30 mT correspond to those of pyrrhotite (Sato et al., 2022), the unstable behavior of the NRM probably arose from the small number of large pyrrhotite crystal grains contained in the Ryugu particles of several hundred micrometers. Coarse-grained pyrrhotite would relax and be demagnetized even if it originally carried stable components. Meanwhile, higher coercivity pyrrhotite may have acquired stable components, but for the reasons mentioned above, such components would be masked by the unstable behavior and thus difficult to detect during AFD treatment. It has also been suggested that magnetite and pyrrhotite crystallized at different times (Tsuchiyama et al., 2024), which may further influence the presence or absence of stable NRM components.

The intensity of the stable NRM component was calculated as $|J_{NRM}(B_{AFD}) - J_{NRM}(B_{AFD-max})|$, where $J_{NRM}(B_{AFD})$ and $J_{NRM}(B_{AFD-max})$ are the NRM vectors at an AFD level of $B_{AFD}$ and at the maximum AF level showing a stable component, respectively. The intensity of the stable NRM component was compared with that of the IRM to estimate IRM-based paleointensity value (Figures 3 and 4). Note that the uncertainty in paleointensity value is calculated from the uncertainty in the slope of the NRM-IRM diagram. The TRM and IRM remanence ratios of the Ryugu samples changed from ranges below to those above 4 mT owing to the difference in remanence behaviors between coarse-grained and framboidal magnetites (Sato et al., 2022). The paleointensity values were subsequently calculated from the NRM-IRM diagram above 4 mT. The IRM-based paleointensity values were estimated to be in the range of 16.3–174 µT (Table 3; Figures 3 and 4).

To estimate TRM-based paleointensity value, the intensity of the stable NRM component of sample C0023-FC009 was plotted with respect to that of TRM (Figure 5). For TRM-based paleointensity estimation with the AFD treatment, such as the Shaw-type paleointensity method, the effect of thermal alteration on the TRM value was evaluated and corrected using the ratio of ARM before and after TRM heating (Rolph and Shaw, 1985). However, the Ryugu samples showed anomalous behavior in ARM, such as significant angular deviations from the applied field direction and dispersions in the stepwise AFD treatments (Figure 6). This anomalous behavior likely arose from strong magnetostatic interactions (Sato et al., 2022; Kimura et al. 2023). For C0023-FC009, a linear trend with a slope of unity was recognized in the diagram showing the original IRM and the IRM after TRM heating in AF ranges below 14 mT (Sato et al., 2022). Subsequently, paleointensity values were calculated using the slope in the AF range of 4–14 mT in the uncorrected NRM-TRM diagram. The TRM-based paleointensity value was estimated to be 56.9 ± 12.0 µT, consistent with the IRM-based paleointensity value of 52.5 ± 7.3 µT.

To select high-fidelity paleointensity values, the following selection criteria were adopted for the stable NRM components of sixteen particles measured in this study:

(1) For all data points of the AFD segments with stable NRM components and above 4 mT, a linear portion with a correlation coefficient larger than 0.9 should be present in the NRM-IRM diagram.

(2) The linear portion in the NRM-IRM diagram should contain more than four points.

Ten particles passed the above criteria, and high-fidelity paleointensity values between 16.3 and 174 µT were obtained, with a mean and standard deviation of 86.2 ± 52.6 µT (Table 3).

Stable components with different directions from those of the artificial components were recognized in the AF ranges of 10–32.5 mT and 0–22 mT in samples A0026 and C0002-4-f, respectively, although laboratory heating TRM and/or laboratory IRM components were recognized in samples A0026 and C0002-4-f in the THD and low-field AFD steps (Nakamura et al., 2022) (Table S3 in Supporting Information; Figure S3 in Supporting Information). The effect of exceptional THD treatment at 110°C on paleointensity estimations on samples A0026 and C0002-4-f was evaluated using the relationship between the stepwise AFD of the original IRM and that after THD at 110°C (Figure 7). A linear portion with unity slope was identified in the 6–22 mT AF range for sample C0076-FO007, indicating that the one-step THD treatment at 110°C affected the remanences of both magnetite with lower coercivity and magnetite/pyrrhotite with higher coercivity, while the effect on those of framboidal magnetite with intermediate coercivity was not significant. Linear portions satisfying the paleointensity criteria above were identified on the NRM-IRM diagrams for samples A0026 and C0002-4-f in the 10–32.5 and 4–22 mT AF ranges, respectively (Table S3 in Supporting Information; Figure S4 in Supporting Information), and the IRM-based paleointensity values were estimated to be 78.3 ± 5.4 and 168 ± 20 µT for A0026 and C0002-4-f, respectively. Although the data points above 4 mT for sample C0002-4-f satisfied the criteria, the NRM-IRM diagram could be clearly divided into two lines, as confirmed by comparing the Akaike information criterion values between one and two lines (Sato et al., 2019). The paleointensity value obtained using the middle coercivity range of 14–22 mT was estimated to be 63.2 ± 9.9 µT (Table S3 in Supporting Information; Figure S3 in Supporting Information). While the paleointensity values of samples A0026 and C0002-4-f agreed with those of the other samples (16.3 and 174 µT), the effect of exceptional THD treatment was recognized in sample C0002-4-f, and the paleointensity values of samples A0026 and C0002-4-f were thus excluded from the high-fidelity paleointensity estimation and the following discussions.

## 4. Discussion
### *Interparticle variations in NRM components*

Previous NRM studies of Ryugu samples (Nakamura et al., 2022; Maurel et al., 2024; Mansbach et al., 2024) reported differing results based on a limited number of particles. Building on these studies, this study conducted stepwise AFD measurements on twenty-three particles from the AO samples, in terms of magnetic records, which had a similar history to those studied by Maurel et al. (2024) and Mansbach et al. (2024), to provide a more statistically robust evaluation. Nineteen of the twenty-three particles exhibited stable NRM components, six (A0225-05, A0225-10, A0225-15, C0213-04, C0213-05, and C0213-07) of which had two stable NRM components (Figure 2). The AFD segments of the stable components were nearly identical to those reported by Mansbach et al. (2024). In contrast, four particles from the AO samples (A0225-11, A0225-13, C0213-03, and C0213-06) did not exhibit stable NRM components, which is consistent with the results of Maurel et al. (2024). Therefore, as a reflection of the original nature of the NRM, some Ryugu particles exhibited stable NRM components, whereas others did not.

According to the interpretation of Maurel et al. (2024), the Ryugu particles originally had no stable components because of a weak to null magnetic field during the remanence acquisition event, and the origin-trending components of samples A0026 and C0002-4-f were magnetically contaminated components. This study conducted stepwise AFD measurements of the NRM of twenty-three AO samples, which had a similar history to those studied by Maurel et al. (2024) and Mansbach et al. (2024). The nineteen particles in this study and three of those studied by Mansbach et al. (2024) exhibited stable NRM components despite having a similar magnetic history to the particles studied by Maurel et al. (2024). Moreover, the initial analysis and AO samples showed similar NRM characteristics, including paleointensity values, regardless of whether they had undergone scanning electron microscopy and X-ray tomography analyses prior to NRM measurements (Table 3; Figures 1–4). The stepwise AFD data for samples C0005 and A0154-a reported by Maurel et al. (2024) were reanalyzed using the criteria for identifying linear components applied in this study. As a result, two stable components were identified for C0005 in the AF ranges of 0–18 and 18–42 mT (Figure S5 in the Supporting Information). The difference in the identification of stable components mainly arises from the use of the origin-trending criterion—defined by the deviation angle from the origin (DANG; Tauxe and Staudigel, 2004)—by Maurel et al. (2024), which was not employed in this study. Thus, the hypothesis that the Ryugu particles originally had no stable components is conclusively rejected.

According to the interpretation of Mansbach et al. (2024), the stable NRM components of A0397, C0085b, and C0006 were VRM acquired on the spacecraft after sampling and/or from Earth's magnetic field prior to the NRM measurements, and the difference with respect to the results in Maurel et al. (2024) was due to the time spent in a null field (11 months for A0154-a, 20 days for C0005, and 3 days for A0397, C0085b, and C0006). This interpretation requires the assumption that the VRM components relaxed to a demagnetized state during magnetically shielded storage. The observation that the twenty-three particles of the AO samples in this study showed both stable and non-stable NRM behaviors, independent of the time spent in a null field (Table 2; Figure 2), provides evidence against the hypothesis of VRM component relaxation (Mansbach et al., 2024). Upon remanence relaxation treatment, C0023-FC009, which exhibits stable NRM components (Figure 1b), did not show significant relaxation before and after storage in the magnetically shielded case for 2 months (Sato et al., 2022), supporting the rejection of the relaxation hypothesis.

Mansbach et al. (2024) conducted VRM acquisition experiments for nineteen days and estimated the upper limit of the VRM intensity acquired over 4.5 years. The estimated VRM intensities were 45 and 42% of the vector-summed LC and MC components for A0397 and C0085b, respectively. Based on magnetic and microscopic observations, the main ferromagnetic minerals in the Ryugu particles were confirmed to be framboidal magnetite, coarse-grained magnetite, and pyrrhotite (Nakamura et al., 2022; Sato et al. 2022). Because the magnetization relaxation times of coarse-grained magnetite and pyrrhotite are much shorter than that of fine-grained framboidal magnetite, the main remanence carriers of VRM acquired during the nineteen days reported by Mansbach et al. (2024) were coarse-grained magnetite and pyrrhotite, whereas framboidal magnetite carried a small proportion of the total VRM. The VRM intensities acquired by framboidal magnetite, which is the main carrier of the LC and MC components in the NRM, were overestimated by Mansbach et al. (2024). Moreover, it is unlikely that particles A0397 and C0085b have consistently remained in the same orientation with respect to the external magnetic fields in the spacecraft and on Earth. The VRM intensities were expected to be much lower than those estimated by Mansbach et al. (2024). Thus, the VRM hypothesis has difficulty explaining the observed NRM characteristics in terms of the observed remanence intensity.

*Intraparticle variations in NRM components*

Particles A0225-03 and A0225-05 originally constituted a single particle, which was split into two daughter particles prior to NRM measurements (Figures S1 and S2). The NRM components of the particles were not mutually-oriented. A0225-03 exhibited one stable component, whereas A0225-05 exhibited two (Figure 2), indicating that the NRM direction of A0225-05 of its lower- and/or higher-coercivity components differed from that of A0225-03 in the original particle. Thus, the original particle comprising A0225-03 and A0225-05 had NRM of spatially inhomogeneous directions within particle.

The NRM of C0085 was measured using SQUID microscopes at the California Institute of Technology and Kochi University (Kirschvink et al., 2023). Note that the name of sample C0085 was incorrectly reported as C0058 by Kirschvink et al. (2023). Measurements revealed the presence of multiple strong magnetic dipole sources with dispersed orientations, the moments of which were comparable to that of the main dipole source. After these measurements, C0085 particle split into two mutually-oriented particles (C0085a and C0085b), and then, the NRM of two pieces were measured using the SQUID microscope at the California Institute of Technology (Mansbach et al., 2024). The NRM direction of C0085a had an inclination and declination of 82.45 and 151.71°, respectively, whereas that of C0085b had an inclination and declination of −40.00 and 316.86°, respectively (Mansbach et al., 2024). These lines of evidence indicate that C0085 had an NRM of spatially inhomogeneous directions within particle.

The spatially inhomogeneous NRM directions in the C0085 and original particle comprising A0225-03 and A0225-05 constrained the time of NRM acquisition to before the final solidification of the particle. Remanence acquisition events that occurred after particle solidification, such as artificial magnetic contamination (Maurel et al., 2024) and VRM acquisition on the spacecraft and/or from Earth's magnetic field (Mansbach et al., 2024), do not adequately explain the observations. The Ryugu particles consist of brecciated fragments ranging from several hundred micrometers to several millimeters in size (Nakamura et al., 2022). One plausible explanation is that the Ryugu particles acquired their NRM before brecciation and preserved it after particle solidification. However, the relationship between the brecciated domains and the remanence directions within the particle remains to be confirmed.

*Interparticle variations in paleointensity*

Paleointensity values were estimated using the IRM-based paleointensity method for the sixteen Ryugu particles with stable NRM components. High-fidelity paleointensity values were obtained from the ten particles that passed the selection criteria with paleointensity values ranging between 16.3 and 174 µT (Table 3; Figures 3 and 4). Similar paleointensity values in nearly identical demagnetization segments of the ten particles suggest that the stable NRM components of these particles have the same origin. Moreover, one of these particles (C0023-FC009) was accepted for application of the TRM-based paleointensity method and showed an identical paleointensity value to its IRM-based paleointensity value (Figures 3a and 5), indicating that its stable NRM component was remanent magnetization with properties similar to its TRM.

To evaluate the nature of the NRM, Maurel et al. (2024) and Mansbach et al. (2024) used NRM/IRM ratios and paleointensity values. The NRM/IRM ratios of A0026 and C0002-4-f were 1–2 orders of magnitude higher than those of A0154-a, A0154-b, and C0005 (Maurel et al., 2024), and the paleointensity values of A0026 and C0002-4-f were also higher than those of the Ryugu particles studied by Maurel et al. (2024) and Mansbach et al. (2024). According to the interpretation of Maurel et al. (2024) and Mansbach et al. (2024), the high NRM/IRM ratios and paleointensity values were the result of artificial magnetic contamination. However, the high paleointensity values of up to 174 µT were obtained from the Ryugu particles having a similar magnetic history to those studied by Maurel et al. (2024) and Mansbach et al. (2024). Therefore, as a reflection of the original nature of the NRM, the Ryugu particles exhibited both stable and non-stable remanence behaviors with various NRM/IRM ratios and corresponding paleointensity values.

Gattacceca and Rochette (2004) evaluated a factor-of-two uncertainty for the REM′ paleointensity method, which is based on derivatives of multiple data points. In contrast, the IRM-based paleointensity estimates in this study are derived from the slope obtained by linear regression of multiple data points. Although these approaches are not strictly equivalent, the IRM-based paleointensity method used in this study is expected to involve uncertainties on the order of a factor of two. For the Ryugu samples, two particles (A0064-FO018 and C0023-FC009) yielded paleointensity constants of 2584 and 4053 µT, respectively, when the IRM-based paleointensity method was applied (Sato et al., 2022). The difference between these values corresponds to a variation on the order of a factor of two and is therefore broadly consistent with the uncertainty range expected for the REM′ method. Weak paleointensity values such as 16.3 ± 2.2 µT (A0225-02) and 21.7 ± 1.5 µT (A0225-03) fall outside the factor-two uncertainty. NRM acquisition before brecciation naturally explains the paleointensity variation among the particles. The remanence vectors of brecciated domains were randomly distributed in one particle, thus weakening the overall NRM. If a particle contains brecciated domains with similar remanence intensities but different directions, the paleointensity value estimated from the particle should be smaller than the true paleointensity value. It is important to confirm the relationship between brecciated domain structures and paleointensity values in future studies.

***Origin of the paleomagnetic record in Ryugu particles***

The NRM characteristics of Ryugu particles obtained in this study, as well as those obtained by Maurel et al. (2024) and Mansbach et al. (2024), are summarized below. Although the initial analysis and AO samples exhibited similar NRM characteristics, regardless of whether they underwent prior scanning electron microscopy and X-ray tomography analyses, only the results from the AO samples—which were not subjected to any other measurements before NRM measurements with stepwise AFD treatments—are included in the summary. The summary includes the results from AO samples with stepwise AFD treatments from this study ($N = 23$), along with those from Maurel et al. (2024) (C0005 and A0154-a) and Mansbach et al. (2024) (A0397, C0085b, and C0006). Note that for C0005, the reanalyzed results are used in the summary.

(1) Twenty-three of the total (twenty-eight particles) exhibited stable NRM components.
(2) Five of the total did not exhibit a stable NRM component.
(3) Eight particles exhibited two stable components.
(4) Two particles exhibited spatially inhomogeneous NRM directions within particle.
(5) IRM-based paleointensity values obtained from stable NRM components vary from 3.15 (Mansbach et al., 2024) to 174 µT (this study).

To explain these characteristics, it is necessary to consider the mechanism of NRM acquisition.

The spatially inhomogeneous NRM directions within particle require the acquisition of the NRM before the final solidification of the particle. Acquisition events that occurred after particle solidification, such as VRM and artificial IRM, were excluded.

The first two characteristics, i.e., that the Ryugu particles exhibited both stable and non-stable remanence behaviors, could arise from the following possibilities (Figure 8):

(A) A particle with a stable component contained a sufficient amount of framboidal magnetite grew in a strong external field and retained stable NRM.
(B) The timescale of framboidal magnetite grain growth events in Ryugu's parent body was longer than that of the external magnetic field fluctuations. In possibility A, the framboidal magnetite within the particle formed under a strong field, whereas in other particles lacking a stable component, it formed under weak or null fields (i.e., it was not originally magnetized).
(C) Unstable remanence with respect to AFD treatment, carried by coarse-grained magnetite and pyrrhotite, was prominent and masked the stable NRM carried by framboidal magnetite.
(D) A particle contained many brecciated domains with random NRM directions and comparable intensities, such that the remanence vectors canceled each other out across the particle as a whole.

One or a combination of these possibilities may explain the observed NRM characteristics. Combination of possibilities A and D could explain the spatially inhomogeneous NRM directions within particle and large variation in paleointensity values. If a Ryugu particle comprises two main brecciated domains with different remanence directions and coercivities, the particle could have two stable NRM components. In possibility B, during the transition of the external magnetic field from a strong to a weak state, some framboidal magnetites may have recorded intermediate field intensities, which could partly explain the variation in estimated paleointensity values among the particles.

Most of the stable NRM components in the Ryugu particles are not origin-trending, and their remanence vectors often show unstable behavior above critical AF levels of 10–30 mT. This instability likely resulted from the small number of large pyrrhotite grains. In this case, the stable NRM component of framboidal magnetite trends toward the total remanence of coarse-grained pyrrhotites, whose intensity is comparable to or greater than that of the stable NRM carried by framboidal magnetite. Consequently, once the coercivity of the coarse-grained pyrrhotite is exceeded, its total remanence changes randomly with large amplitudes during each step of AFD treatment. Similar behavior during stepwise AFD treatment has been reported in single zircon crystals containing magnetite and pyrrhotite (Sato et al., 2015), supporting this interpretation. Under these circumstances, the origin-trending criterion is not appropriate for evaluating whether the magnetization preserves a primary magnetic record. Regarding the detectability of stable NRM components, an increase in Ryugu particle size likely involves a trade-off between a greater number of coarse-grained pyrrhotite crystals within each particle and a greater number of brecciated fragments it contains. Future studies should examine the relationship between the states of brecciated fragments and the NRM characteristics to gain further insight into the nature of the NRM records.

The most likely acquisition mechanism of stable NRM components is chemical remanent magnetization (CRM) associated with the growth of framboidal magnetite. Because the brecciation events occurred after an aqueous alteration event in Ryugu's parent body (Yamaguchi et al., 2023), remanence acquisition by CRM is consistent with the requirement that the NRM should have been acquired before particle solidification. Dolomite grains in the Ryugu particles often contain small crystals of magnetite (Nakamura et al., 2022), and both probably precipitated from the same aqueous fluid because of their similar oxygen isotope compositions (Yokoyama et al., 2023). The time of dolomite precipitation was determined using Mn–Cr dating, revealing that it occurred 3.1–6.8 Myr after the formation of calcium-aluminum-rich inclusions (CAIs) (Yokoyama et al., 2023). The Mn–Cr dating of carbonate in different Ryugu particles revealed an older age of <1.8 Myrs after CAI formation (McCain et al., 2023), which resulted from the use of matrix-matched standards as opposed to calcite standards used exclusively in previous studies on carbonaceous chondrites. We proceeded to use the younger age of 3.1–6.8 Myr after CAI formation. These lines of evidence suggest that magnetite crystallized and grew simultaneously with dolomite 3.1–6.8 Myr after CAI formation. Subsequent temperature increases after the aqueous alteration event were not as high as the daytime temperature of the current Ryugu surface ($\leq 100°C$)—at least not for the returned samples (Nakamura et al., 2022; Okazaki et al., 2023; Yokoyama et al., 2023). Features indicative of strong deformation or shock melting were not observed in the Ryugu samples, indicating that they did not experience intense shock (Nakamura et al., 2022). Therefore, the stable NRM components carried by the framboidal magnetites were unlikely to be TRM or shock remanent magnetization acquired during a later catastrophic disruption event, but were attributed to CRM acquired during aqueous alteration in Ryugu's parent body.

An important remaining issue is the large variation in paleointensity values. The IRM-based paleointensity estimates for the Ryugu samples range from 3.15 to 174 µT. The IRM-based method intrinsically carries a factor-of-two uncertainty when the NRM is of TRM origin (Gattacceca and Rochette, 2004). However, if the NRM is of CRM origin, the paleointensity derived from the IRM-based method may deviate from the actual field intensity by the CRM/TRM ratio. For example, lunar samples have shown order-of-magnitude differences between IRM-based and TRM-based paleointensity estimates (Tarduno et al., 2021; Zhou et al., 2024). In particular, the CRM/TRM ratio for framboidal magnetite needs to be quantitatively constrained in future studies. Conservatively, the actual paleointensity is currently estimated to lie within a range of several to a few hundred microteslas.

Considering the CRM acquisition ages of 3.1–6.8 Myr after CAI formation, these ages may span the transition from the presence to the dispersal of the solar nebula (e.g., Wang et al., 2017). After the clearing of the nebula, solar wind-related magnetic fields could have magnetized the particles through direct interaction and field amplification (O'Brien et al., 2020; Anand et al., 2022). Although such fields are generally estimated to have strengths of only a few microteslas (O'Brien et al., 2020), they could still plausibly account for the observed NRM if the actual paleointensity was of similar magnitude. Alternatively, the NRM may record a nebular field acquired either in a distant region of the disk or during the late stage of disk evolution.

Notably, a TRM-based paleointensity, one of the most reliable paleointensity approaches, was successfully obtained for sample C0023-FC009. The TRM-based paleointensity value (56.9 ± 12.0 µT) is consistent with the IRM-based estimate (52.5 ± 7.3 µT), and also broadly consistent with the average IRM-based value of 86.2 ± 52.6 µT obtained in this study. Assuming that the CRM/TRM ratio does not significantly deviate from unity, the actual paleointensity is likely on the order of several tens of microteslas, suggesting that the magnetization may record a nebular field acquired in the inner region of the protoplanetary disk.

Accurate determination of paleointensity is therefore crucial for constraining the magnetic environment of the early solar system. Future studies should aim to obtain TRM-based paleointensity estimates from additional Ryugu samples, while carefully considering the NRM characteristics of the particles, to better elucidate the nature of the external magnetic field that existed during CRM acquisition.

The NRM characteristics obtained in this study also provide important insights for magnetic field investigations of asteroids. The magnetometer on the MASCOT lander measured the surface magnetic field of asteroid Ryugu (Hercik et al., 2020). No magnetic field above the instrument's detection limit was observed at a distance of 10–20 cm from the surface, indicating that the surface material was not unidirectionally magnetized on decimeter scales exceeding magnetization values of $\sim 10^{-5}$ Am$^2$/kg (Hercik et al. 2020). This observation is consistent with the NRM characteristics: (1) millimeter-scale particles exhibited magnetization values of $3-19 \times 10^{-4}$ Am$^2$/kg (Maurel et al., 2024; Mansbach et al., 2024), and (2) two particles with millimeter scale showed heterogeneous magnetization. The unidirectionally magnetized region is likely smaller than the size of brecciated domains. Considering the brecciation events that occurred after the acquisition of the external magnetic field, a precise evaluation of the magnetic environment from magnetometer observations on asteroids should ideally be based on measurements taken at distances comparable to the brecciation scale from the surface material.

## 5. Conclusion

This study conducted stepwise AFD measurements of NRM on twenty-eight Ryugu particles and IRM-based paleointensity measurements on sixteen of these particles with stable NRM components to precisely understand the characteristics of NRM records in Ryugu particles. The following characteristics were obtained from these measurements: (1) Twenty-three out of the twenty-eight particles exhibited stable NRM components, while the remaining five particles did not. (2) One particle that was split into two daughter samples exhibited spatially inhomogeneous NRM directions within particle. (3) IRM-based paleointensity values obtained from stable NRM components varied from 16.3 to 174 µT. These NRM characteristics are consistent with those observed in previous studies. The spatially inhomogeneous NRM directions place the constraint that the observed remanence should have been acquired before the final solidification of the current Ryugu particles. Remanence acquisition events after particle solidification, such as artificial IRM and VRM on the spacecraft and Earth, are inconsistent with the observed characteristics. The remanence acquisition mechanism that explains the observed NRM characteristics is CRM associated with the growth of framboidal magnetite during aqueous alteration of Ryugu's parent body. Such magnetizations may record ambient magnetic fields of the solar nebula or the solar wind acquired during different stages of magnetite growth.

**Conflict of Interest Statement**

The authors declare no conflicts of interest relevant to this study.

**Data availability Statement**

Data from this paper are archived at the UTokyo Repository (Sato, 2025).

**Acknowledgements**

The authors thank the two anonymous reviewers for their thoughtful and thorough reviews. MS is supported by JSPS KAKENHI grant JP21H01140. MS is supported by the Hypervelocity Impact Facility (former name: The Space Plasma Laboratory), Institute of Space and Astronautical Science (ISAS), Japan Aerospace Exploration Agency (JAXA).

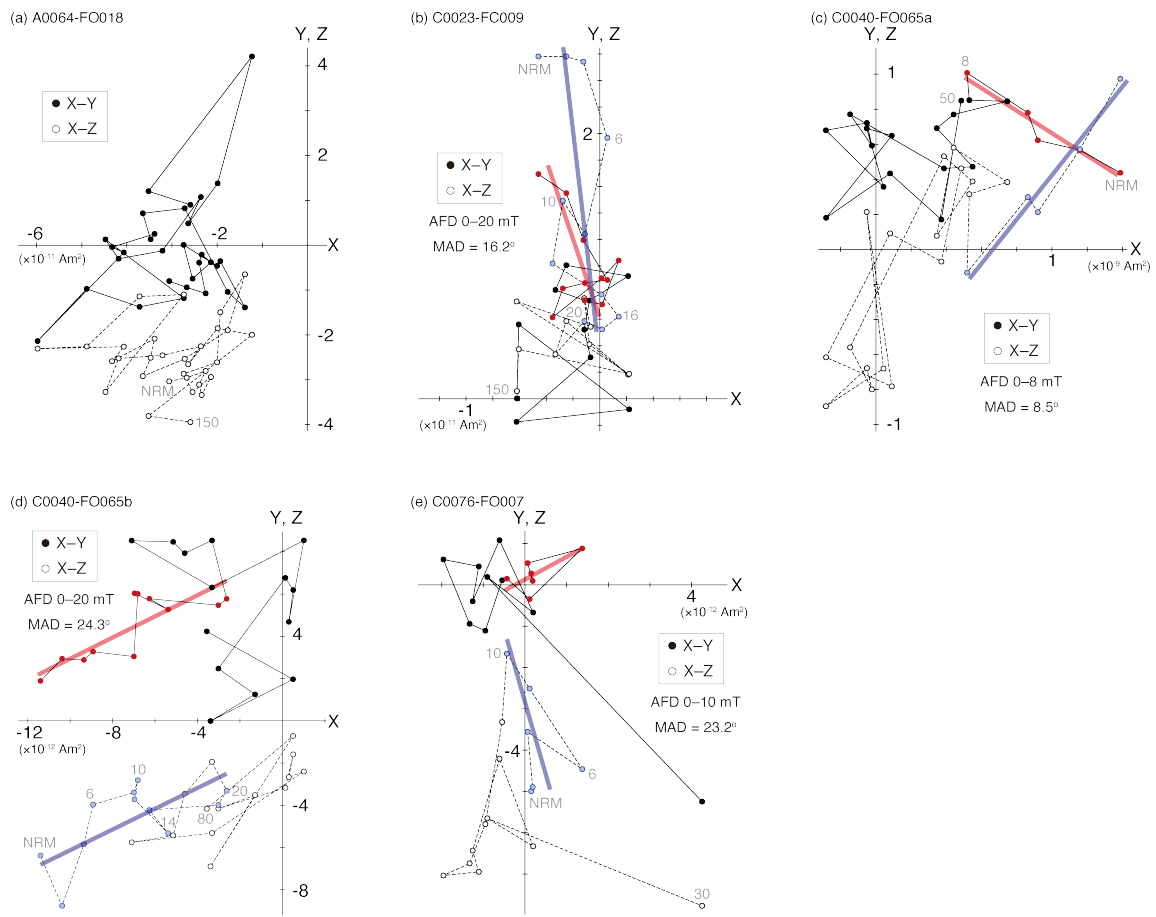

**Figure 1.** Orthogonal vector plots for stepwise alternating field demagnetization of natural remanent magnetization of the initial analysis samples. The closed and open symbols denote the horizontal and vertical projections, respectively.

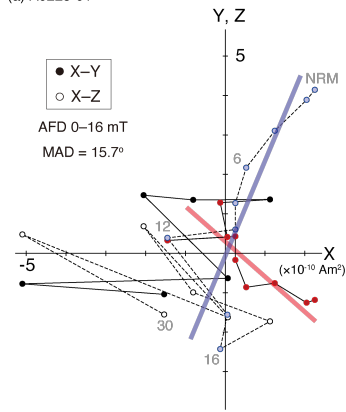
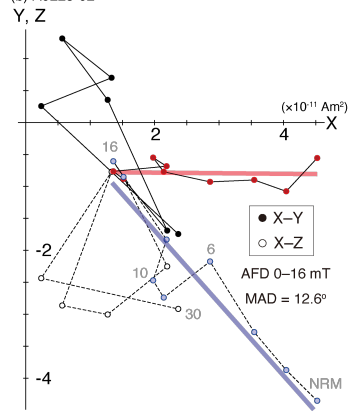
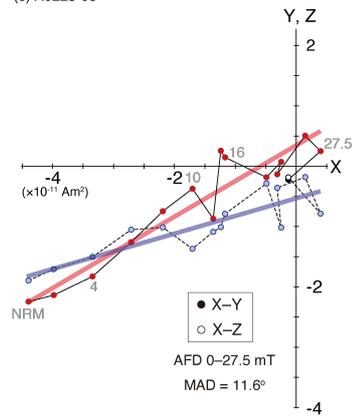
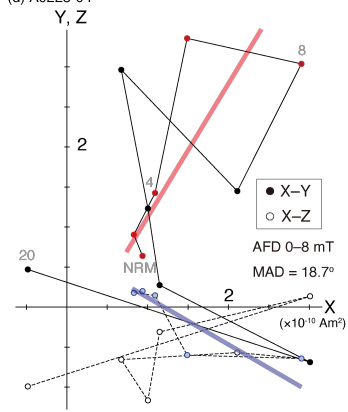
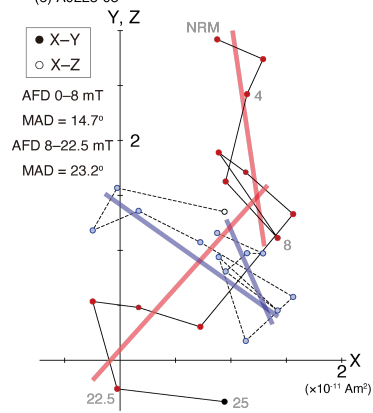
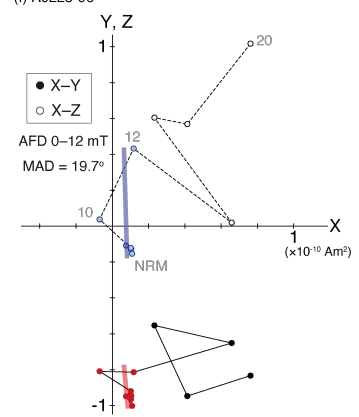
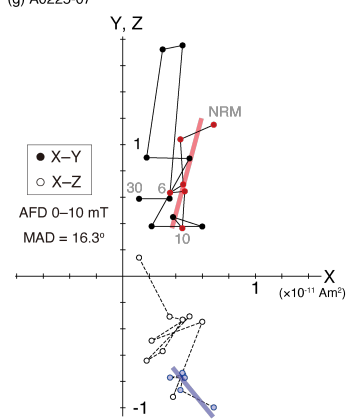
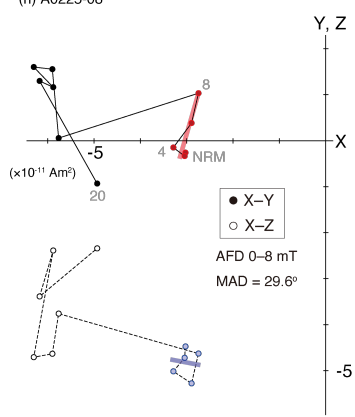
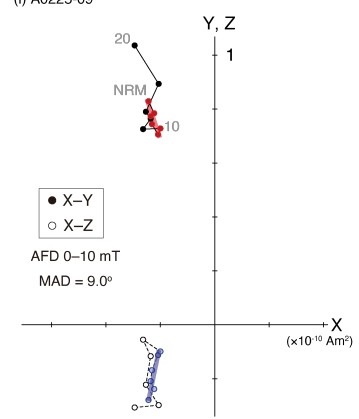

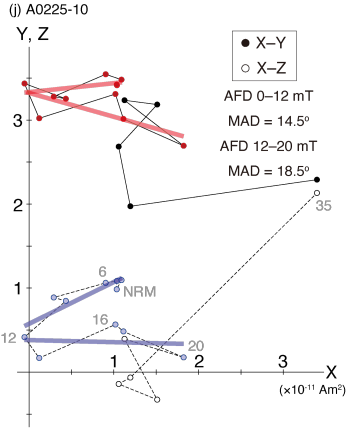
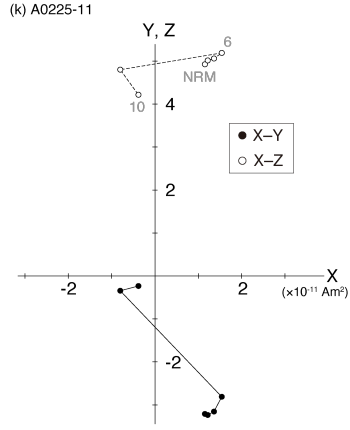
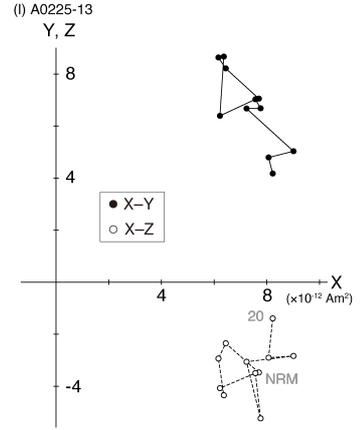
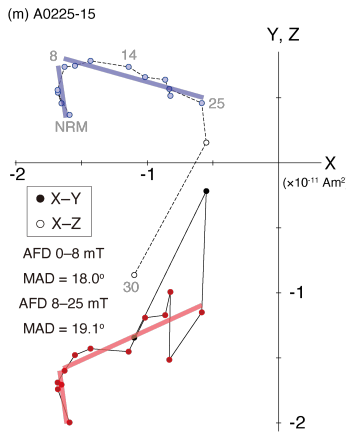
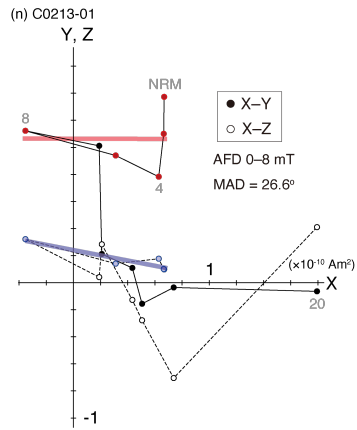
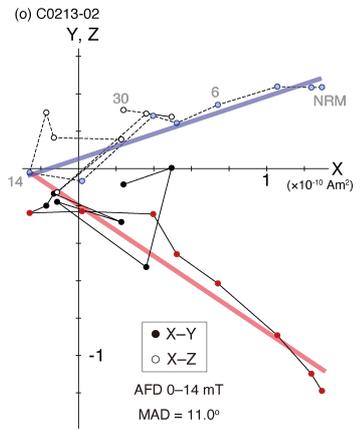
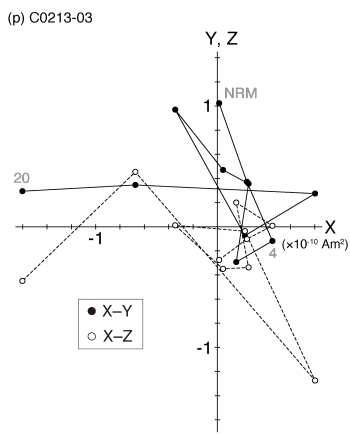
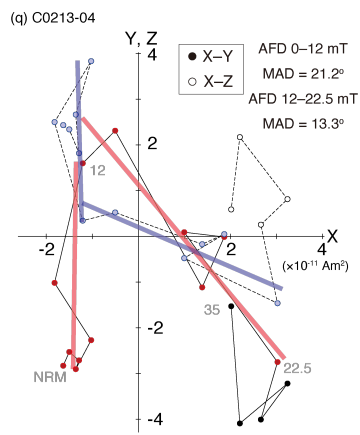
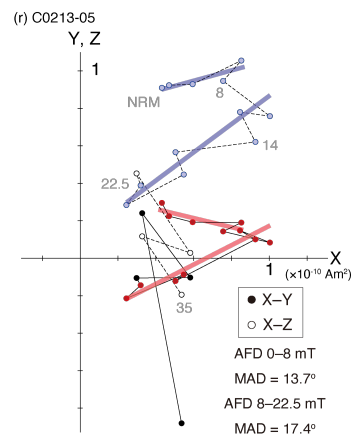

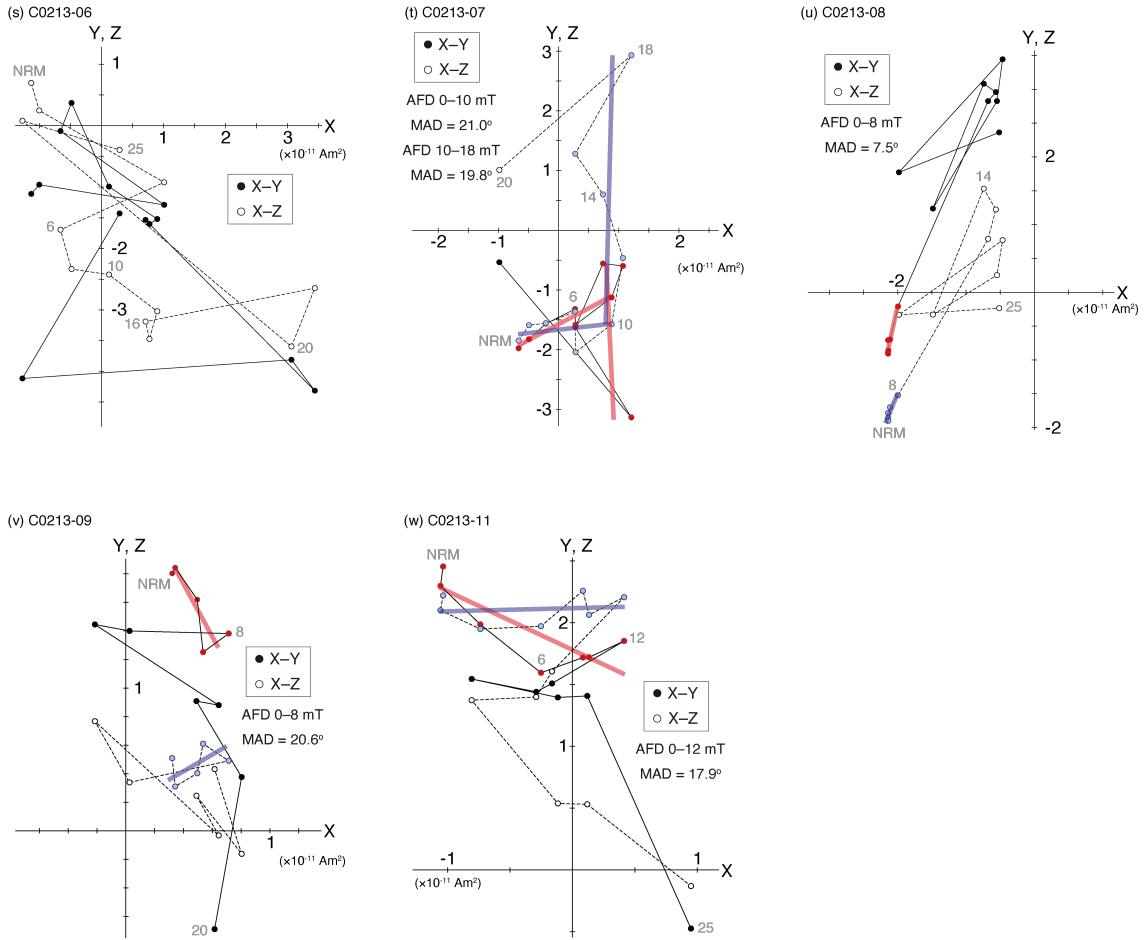

**Figure 2.** Orthogonal vector plots for stepwise alternating field demagnetization of natural remanent magnetization of the AO samples. The closed and open symbols denote the horizontal and vertical projections, respectively.

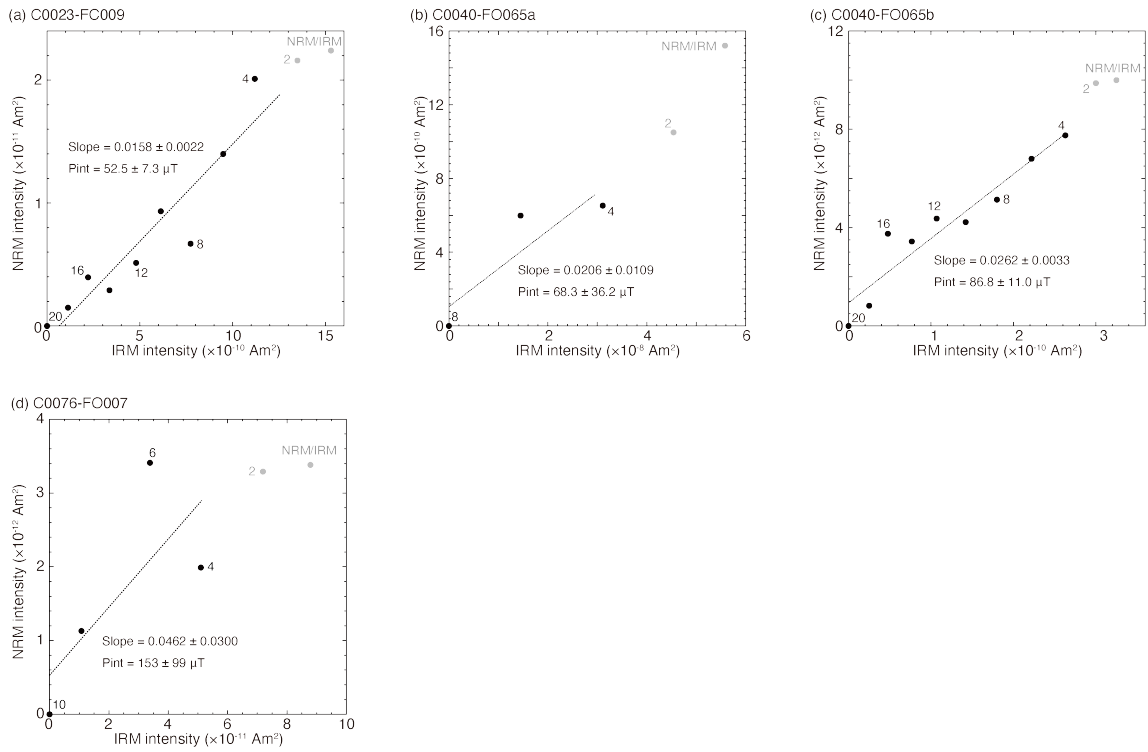

**Figure 3.** The paleointensity estimation for the initial analysis samples. The intensity of the natural remanent magnetization (NRM) is plotted as a function of that of the isothermal remanent magnetization (IRM) for the same alternating field demagnetization steps. Black symbols indicate the linear portion used for the paleointensity calculation.

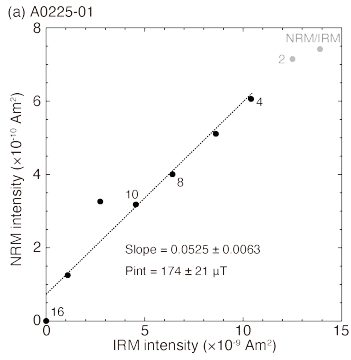
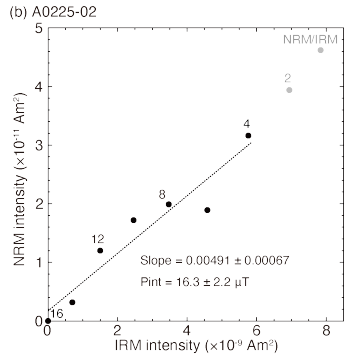
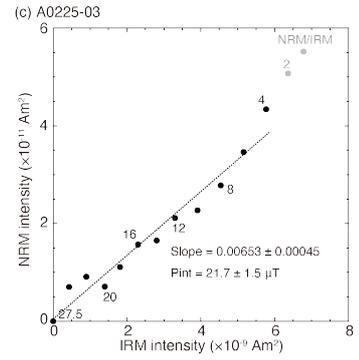
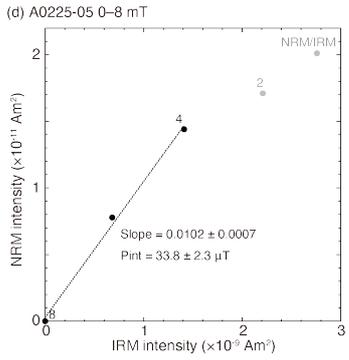
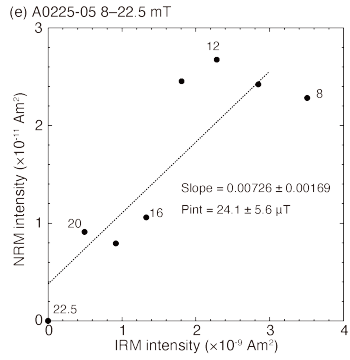
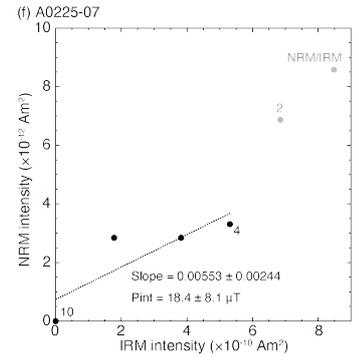
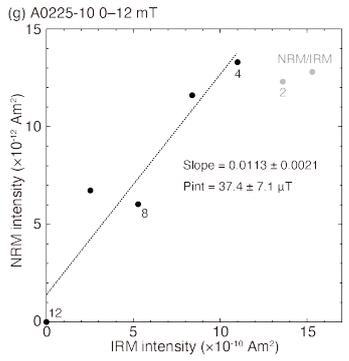
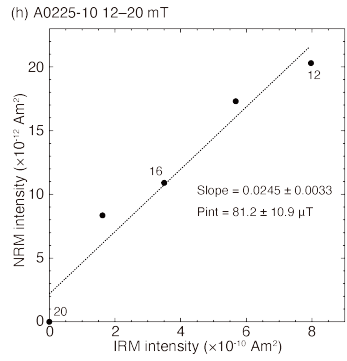
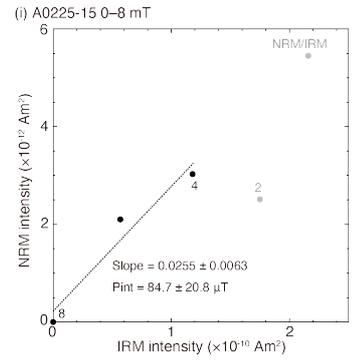

continue

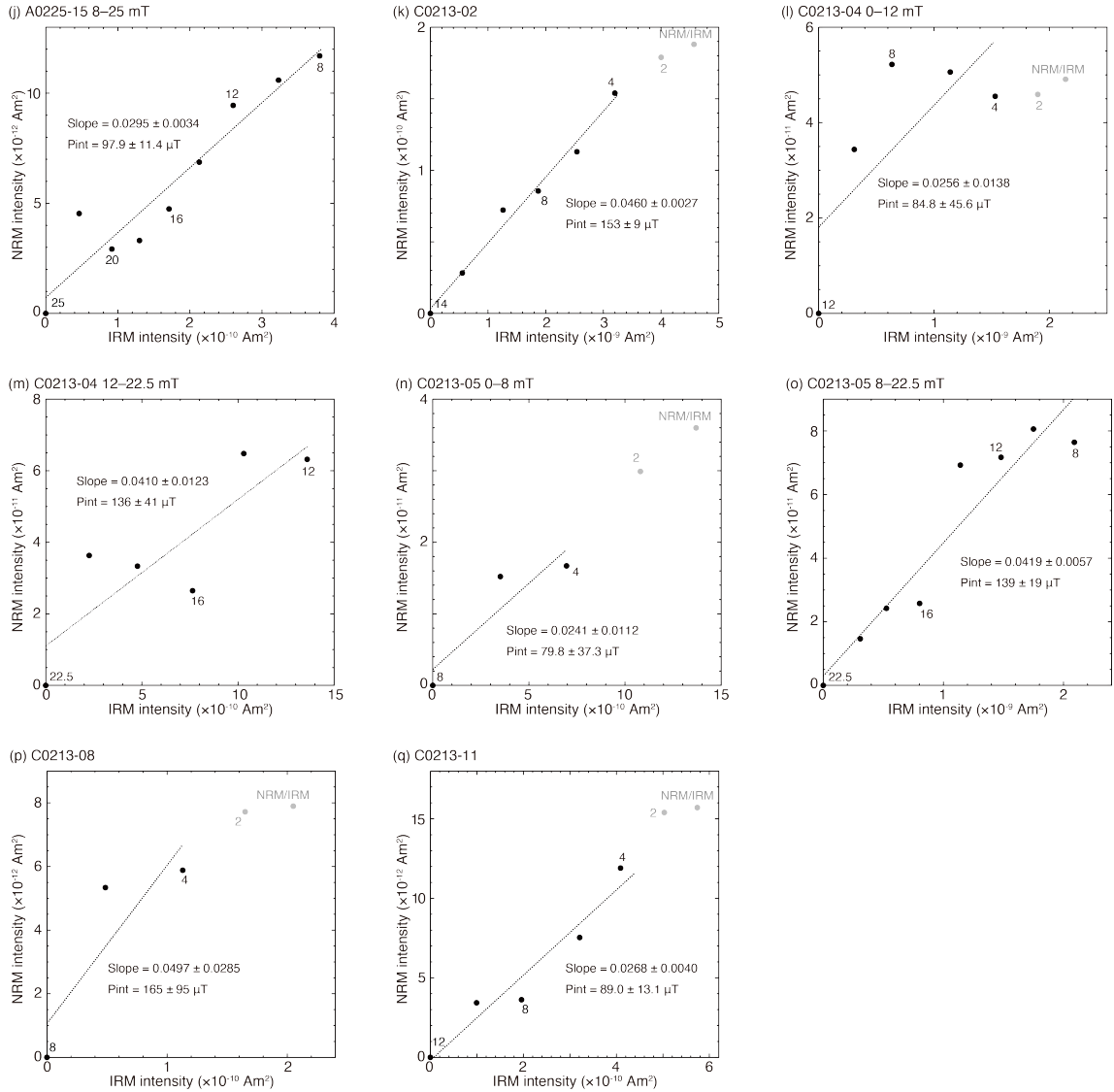

**Figure 4.** The paleointensity estimation for the AO samples. The intensity of the natural remanent magnetization (NRM) is plotted as a function of that of the isothermal remanent magnetization (IRM) for the same alternating field demagnetization steps. Black symbols indicate the linear portion used for the paleointensity calculation.

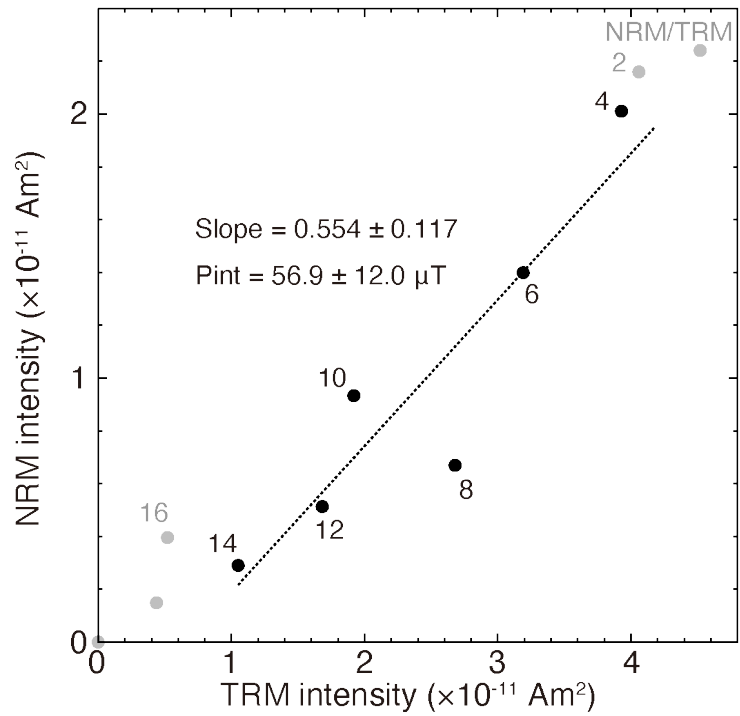

**Figure 5.** Results of the Shaw-type paleointensity experiment for the C0023-FC009 sample. The intensity of the natural remanent magnetization (NRM) is plotted as a function of that of the thermal remanent magnetization (TRM) for the same alternating field demagnetization steps. Black symbols indicate the linear portion used for the paleointensity calculation.

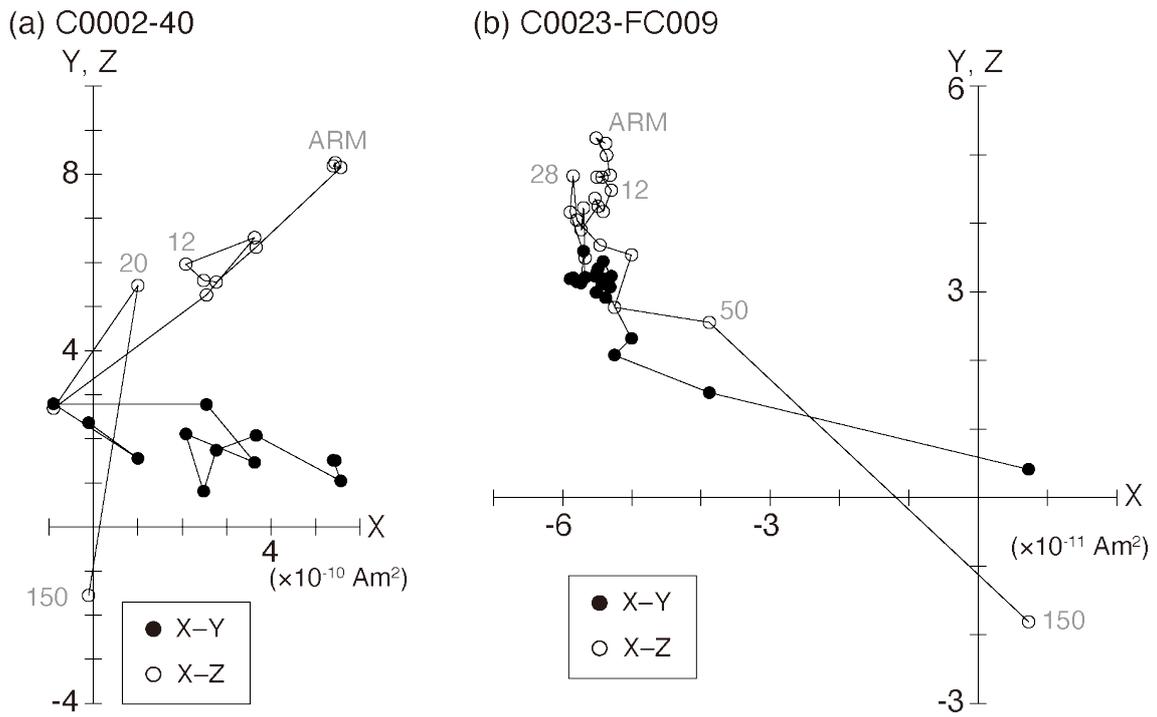

**Figure 6.** Orthogonal vector plots for stepwise alternating field demagnetization of anhysteretic remanent magnetization. Closed and open symbols denote horizontal and vertical projections, respectively.

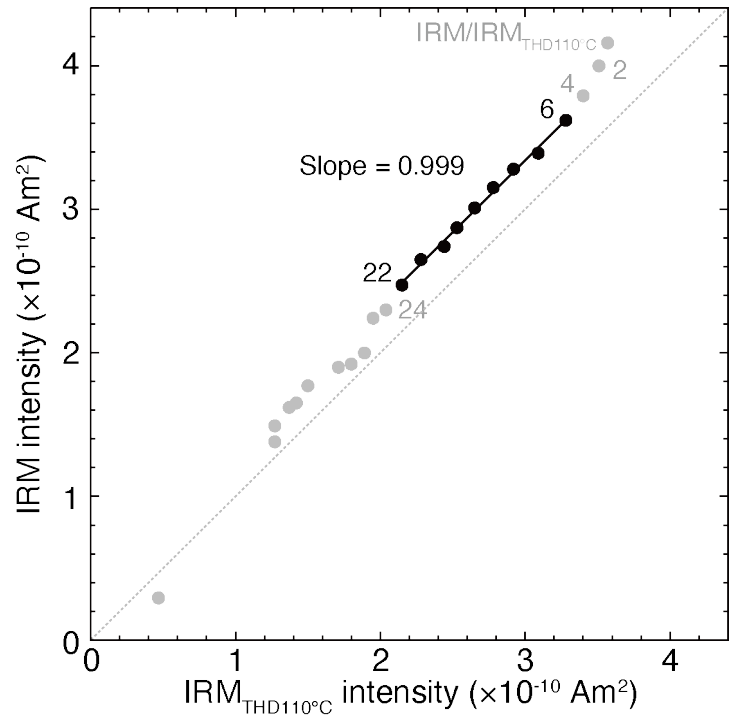

**Figure 7.** The intensity of isothermal remanent magnetization (IRM) was plotted as a function of the intensity after thermal demagnetization (THD) at 110°C for the same alternating field demagnetization steps. Black symbols indicate the linear portion with a unit slope.

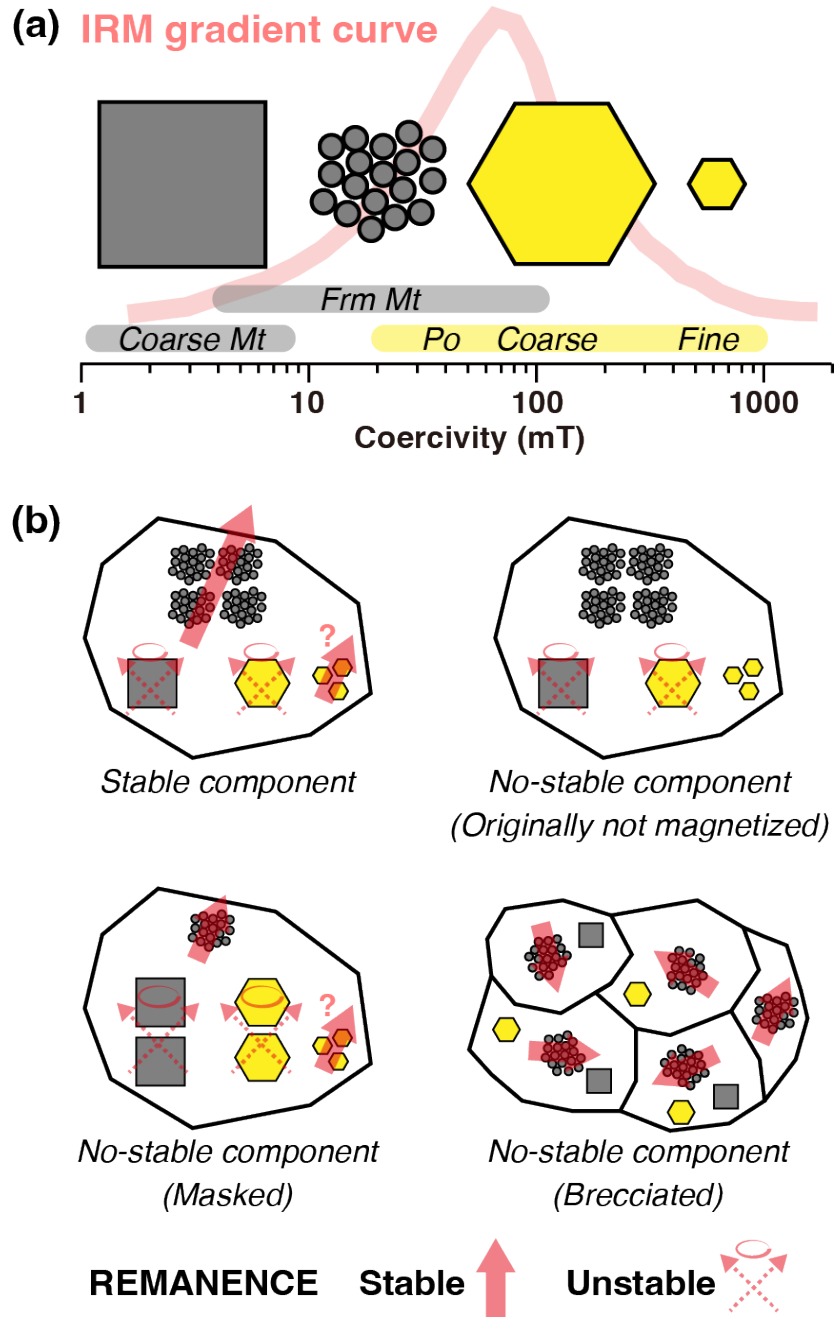

**Figure 8.** Interpretation of magnetic measurements. (a) Typical coercivity ranges of magnetic minerals for coarse grained magnetite, framboidal magnetite, and pyrrhotite. Isothermal remanent magnetization (IRM) gradient curves from Sato et al. (2022). (b) Interpretation of natural remanence records (see text).

**Table 1.** Summary of experimental samples

| Series | Sample | Measurement sequence of this study |
|---|---|---|
| Initial analysis | A0064-FO018 | Stepwise AFD of NRM |
| | C0002-40 | Stepwise AFD of ARM |
| | C0023-FC009 | Stepwise AFD of NRM, ARM, IRM, TRM |
| | C0040-FO065a | Stepwise AFD of NRM and IRM |
| | C0040-FO065b | Stepwise AFD of NRM and IRM |
| | C0076-FO007 | Stepwise AFD of NRM and IRM |
| | | THD and stepwise AFD of IRM |
| AO | A0225-01 | Stepwise AFD of NRM and IRM |
| | A0225-02 | Stepwise AFD of NRM and IRM |
| | A0225-03 | Stepwise AFD of NRM and IRM |
| | A0225-04 | Stepwise AFD of NRM |
| | A0225-05 | Stepwise AFD of NRM and IRM |
| | A0225-06 | Stepwise AFD of NRM |
| | A0225-07 | Stepwise AFD of NRM and IRM |
| | A0225-08 | Stepwise AFD of NRM |
| | A0225-09 | Stepwise AFD of NRM |
| | A0225-10 | Stepwise AFD of NRM and IRM |
| | A0225-11 | Stepwise AFD of NRM |
| | A0225-13 | Stepwise AFD of NRM and IRM |
| | A0225-15 | Stepwise AFD of NRM and IRM |
| | C0213-01 | Stepwise AFD of NRM |
| | C0213-02 | Stepwise AFD of NRM and IRM |
| | C0213-03 | Stepwise AFD of NRM |
| | C0213-04 | Stepwise AFD of NRM and IRM |
| | C0213-05 | Stepwise AFD of NRM and IRM |
| | C0213-06 | Stepwise AFD of NRM |
| | C0213-07 | Stepwise AFD of NRM |
| | C0213-08 | Stepwise AFD of NRM and IRM |
| | C0213-09 | Stepwise AFD of NRM |
| | C0213-11 | Stepwise AFD of NRM and IRM |

THD: thermal demagnetization; AFD: alternating field demagnetization; NRM: natural remanent magnetization; IRM: isothermal remanent magnetization; ARM: anhysteretic remanent magnetization; TRM: thermoremanent magnetization.

**Table 2.** Storage duration in magnetic shield case for the AO samples

| Sample | Duration (Day) |
|---|---|
| A0225-01 | 3 |
| A0225-02 | 3 |
| A0225-03 | 4 |
| A0225-04 | 4 |
| A0225-05 | 5 |
| A0225-06 | 5 |
| A0225-07 | 6 |
| A0225-08 | 9 |
| A0225-09 | 9 |
| A0225-10 | 9 |
| A0225-11 | 12 |
| A0225-13 | 12 |
| A0225-15 | 12 |
| C0213-01 | 3 |
| C0213-02 | 3 |
| C0213-03 | 4 |
| C0213-04 | 4 |
| C0213-05 | 6 |
| C0213-06 | 6 |
| C0213-07 | 7 |
| C0213-08 | 9 |
| C0213-09 | 10 |
| C0213-11 | 10 |

**Table 3.** Summary of paleomagnetic and paleointensity experiments

| Sample | Linear segment (mT) | MAD (Degree) | AF segment (mT) | N | Slope | R | Palaeointensity (μT) |
|---|---|---|---|---|---|---|---|
| A0064-FO018 | N/A | N/A | N/A | N/A | N/A | N/A | N/A |
| C0023-FC009 | 0–20 | 16.2 | 4–20 | 9 | 0.0158 ± 0.0022 | 0.938 | 52.5 ± 7.3 |
| C0040-FO065a | 0–8 | 8.5 | 4–8 | (3) | 0.0206 ± 0.0109 | (0.883) | (68.3 ± 36.2) |
| C0040-FO065b | 0–20 | 24.3 | 4–20 | 9 | 0.0262 ± 0.0033 | 0.948 | 86.8 ±11.0 |
| C0076-FO007 | 0–10 | 23.2 | 4–10 | (4) | 0.0462 ± 0.0300 | (0.737) | (153 ± 99) |
| A0225-01 | 0–16 | 15.7 | 4–16 | 7 | 0.0525 ± 0.0063 | 0.966 | 174 ± 21 |
| A0225-02 | 0–16 | 12.6 | 4–16 | 7 | 0.00491 ± 0.00067 | 0.957 | 16.3 ± 2.2 |
| A0225-03 | 0–27.5 | 11.6 | 4–27.5 | 12 | 0.00653 ± 0.00045 | 0.977 | 21.7 ± 1.5 |
| A0225-04 | 0–8 | 18.7 | N/A | N/A | N/A | N/A | N/A |
| A0225-05 | 0–8 | 14.7 | 4–8 | (3) | 0.0102 ± 0.0007 | 0.998 | (33.8 ± 2.3) |
|  | 8–22.5 | 23.2 | 8–22.5 | 8 | 0.00726 ± 0.00169 | (0.869) | (24.1 ± 5.6) |
| A0225-06 | 0–12 | 19.7 | N/A | N/A | N/A | N/A | N/A |
| A0225-07 | 0–10 | 16.3 | 4–10 | (4) | 0.00553 ± 0.00244 | (0.849) | (18.4 ± 8.1) |
| A0225-08 | 0–8 | 29.6 | N/A | N/A | N/A | N/A | N/A |
| A0225-09 | 0–10 | 9.0 | N/A | N/A | N/A | N/A | N/A |
| A0225-10 | 0–12 | 14.5 | 4–12 | 5 | 0.0113 ± 0.0021 | 0.950 | 37.4 ± 7.1 |
|  | 12–20 | 18.5 | 12–20 | 5 | 0.0245 ± 0.0033 | 0.974 | 81.2 ± 10.9 |
| A0225-11 | N/A | N/A | N/A | N/A | N/A | N/A | N/A |
| A0225-13 | N/A | N/A | N/A | N/A | N/A | N/A | N/A |
| A0225-15 | 0–8 | 18.0 | 4–8 | (3) | 0.0255 ± 0.0063 | 0.971 | (84.7 ± 20.8) |
|  | 8–25 | 19.1 | 8–25 | 9 | 0.0295 ± 0.0034 | 0.955 | 97.9 ± 11.4 |
| C0213-01 | 0–8 | 26.6 | N/A | N/A | N/A | N/A | N/A |
| C0213-02 | 0–14 | 11.0 | 4–14 | 6 | 0.0460 ± 0.0027 | 0.993 | 153 ± 9 |
| C0213-03 | N/A | N/A | N/A | N/A | N/A | N/A | N/A |
| C0213-04 | 0–12 | 21.2 | 4–12 | 5 | 0.0256 ± 0.0138 | (0.732) | (84.8 ± 45.6) |
|  | 12–22.5 | 13.3 | 12–22.5 | 6 | 0.0410 ± 0.0123 | (0.857) | (136 ± 41) |
| C0213-05 | 0–8 | 13.7 | 4–8 | (3) | 0.0241 ± 0.0112 | 0.906 | (79.8 ± 37.3) |
|  | 8–22.5 | 17.4 | 8–22.5 | 8 | 0.0419 ± 0.0057 | 0.948 | 139 ± 19 |
| C0213-06 | N/A | N/A | N/A | N/A | N/A | N/A | N/A |
| C0213-07 | 0–10 | 21.0 | N/A | N/A | N/A | N/A | N/A |
|  | 10–18 | 19.8 | N/A | N/A | N/A | N/A | N/A |
| C0213-08 | 0–8 | 7.5 | 4–8 | (3) | 0.0497 ± 0.0285 | (0.867) | (165 ± 95) |
| C0213-09 | 0–8 | 20.6 | N/A | N/A | N/A | N/A | N/A |
| C0213-11 | 0–12 | 17.9 | 4–12 | 5 | 0.0268 ± 0.0040 | 0.969 | 89.0 ± 13.1 |
| | | | | | | | Ave. 86.2 |
| | | | | | | | Std. 52.6 |

Values in parentheses do not fulfill the selection criteria.

Supporting Information for

**[Characteristics of natural remanence records in fine-grained particles returned from asteroid Ryugu]**

[Masahiko Sato[1,2,3]*, Yuki Kimura[4], Tadahiro Hatakeyama[5], Tomoki Nakamura[6], Satoshi Okuzumi[7], Sei-ichiro Watanabe[8], Seiji Sugita[2], Satoshi Tanaka[3], Shogo Tachibana[2,3], Hisayoshi Yurimoto[9], Takaaki Noguchi[10], Ryuji Okazaki[11], Hikaru Yabuta[12], Hiroshi Naraoka[11], Kanako Sakamoto[3], Toru Yada[3], Masahiro Nishimura[3], Aiko Nakato[13], Akiko Miyazaki[3], Kasumi Yogata[3], Masanao Abe[3], Tatsuaki Okada[3], Tomohiro Usui[3], Makoto Yoshikawa[3], Takanao Saiki[3], Fuyuto Terui[14], Satoru Nakazawa[3], and Yuichi Tsuda[3]]

[[1]Department of Physics, Tokyo University of Science, Tokyo 162-8601, Japan. [2]Department of Earth and Planetary Science, the University of Tokyo, Tokyo 113-0033, Japan. [3]Institute of Space and Astronautical Science (ISAS), Japan Aerospace Exploration Agency (JAXA), Sagamihara 252-5210, Japan. [4]Institute of Low Temperature Science, Hokkaido University, Sapporo 060-0819, Japan. [5]Institute of Frontier Science and Technology, Okayama University of Science, Okayama 700-0005, Japan. [6]Department of Earth Sciences, Tohoku University, Sendai 980-8578, Japan. [7]Department of Earth and Planetary Sciences, Institute of Science Tokyo, Tokyo 152-8550, Japan. [8]Department of Earth and Environmental Sciences, Nagoya University, Nagoya 464-8601, Japan. [9]Department of Natural History Sciences, Hokkaido University, Sapporo 001-0021, Japan. [10]Division of Earth and Planetary Sciences, Kyoto University, Kyoto 606-8502, Japan. [11]Department of Earth and Planetary Sciences, Kyushu University, Fukuoka 819-0395, Japan. [12]Graduate School of Advanced Science and Engineering, Hiroshima University, Higashi-Hiroshima 739-8526, Japan. [13]National Institute of Polar Research, Tachikawa 190-8518, Japan. [14]Department of Mechanical Engineering, Kanagawa Institute of Technology, Atsugi 243-0292, Japan.]

**Contents of this file**

- Figures S1–S5
- Tables S1–S3

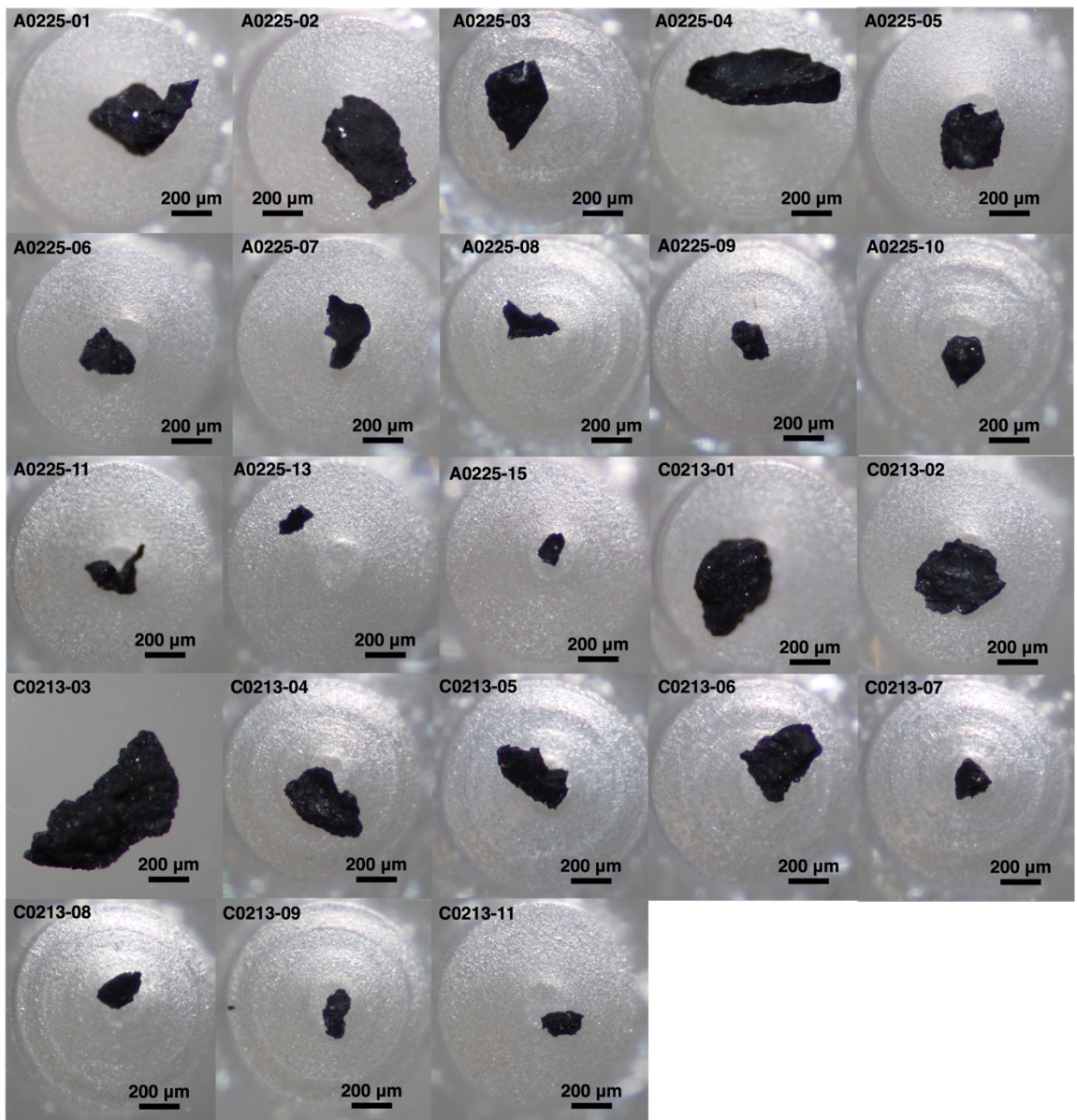

**Figure S1.** Photo images of the AO samples.

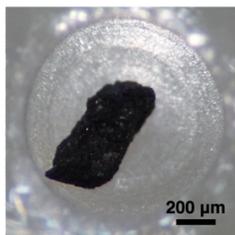

**Figure S2.** Photo images of the original particle of A0225-03 and A0225-05.

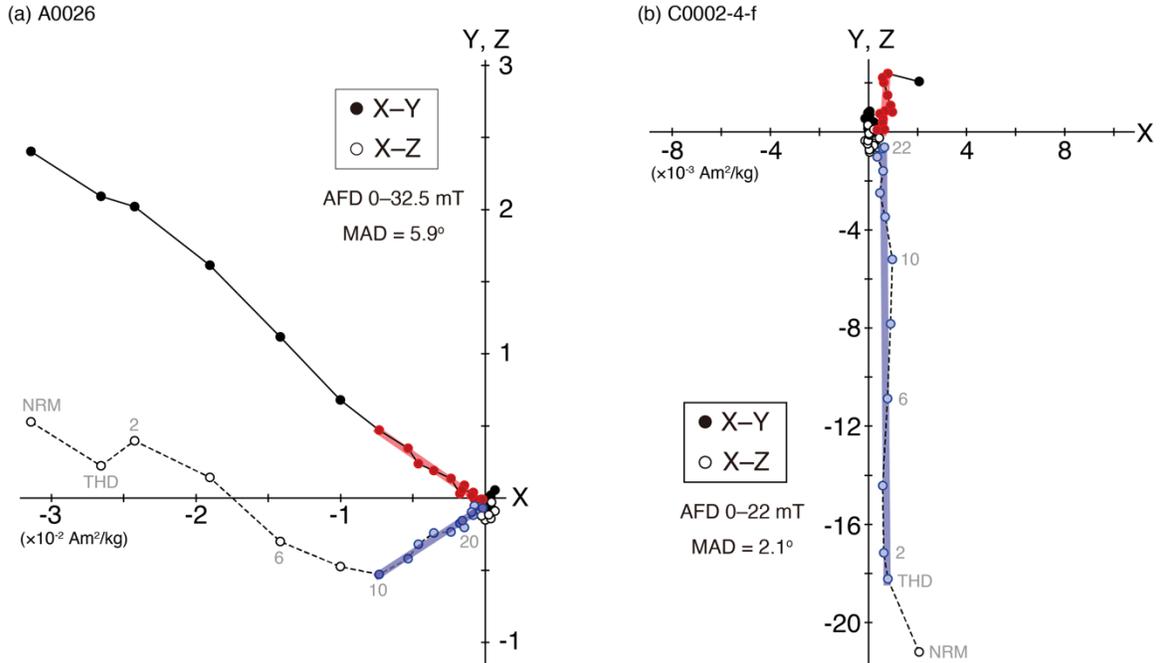

**Figure S3.** Orthogonal vector plots for stepwise alternating field demagnetization of natural remanent magnetization of the initial analysis samples. The closed and open symbols denote the horizontal and vertical projections, respectively.

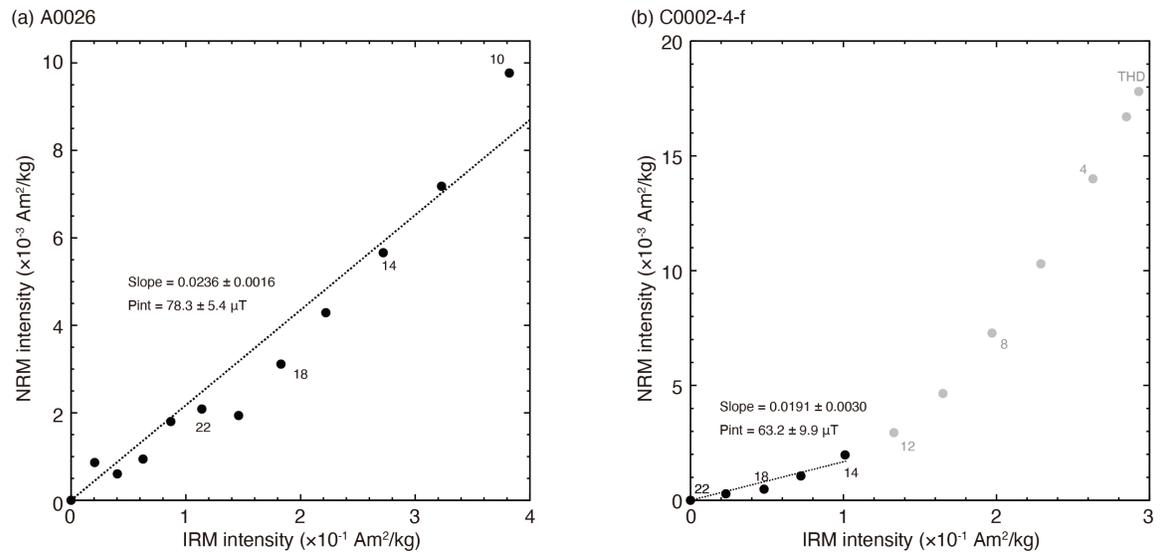

**Figure S4.** The paleointensity estimation for the initial analysis samples. The intensity of the natural remanent magnetization (NRM) is plotted as a function of that of the isothermal remanent magnetization (IRM) for the same alternating field demagnetization steps. Black symbols indicate the linear portion used for the paleointensity calculation.

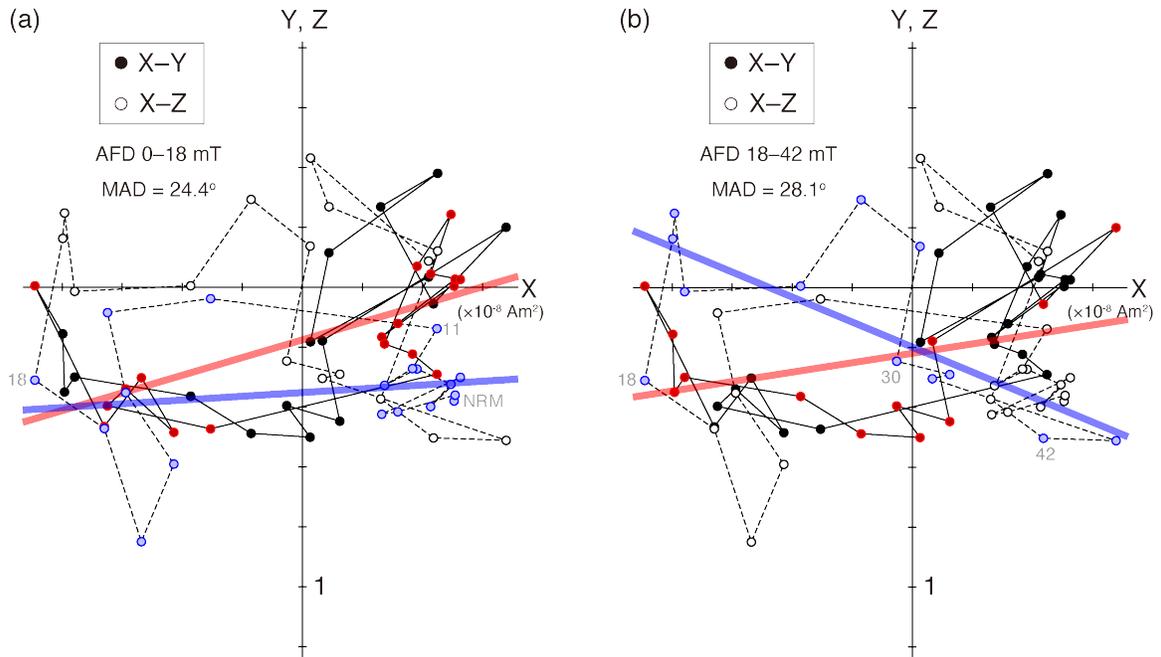

**Figure S5.** Orthogonal vector plots for stepwise alternating field demagnetization of natural remanent magnetization of C0005. The closed and open symbols denote the horizontal and vertical projections, respectively. For clarity, the data are plotted up to 60 mT. (a) 0–18 mT component and (b) 18–42 mT component.

**Table S1.** Summary of potential contaminations

| Source | Magnetic field | Event/Context | Remanence | Sample |
|---|---|---|---|---|
| Ion engine system | Several tens of μT | Transport from Ryugu to Earth | VRM | All |
| Geomagnetic field | Several tens of μT | On Earth | VRM | All |
| Electron microscopy | $10^{-5}$–$10^{0}$ T | Initial analysis | IRM | A0064-FO018 |
| | | | | C0002-40 |
| | | | | C0023-FC009 |
| | | | | C0076-FO007 |
| Heat capacity measurement (100°C) | Several tens of μT | Initial analysis | TRM | A0026 |
| Glue treatment (80°C) | Several tens of μT | Initial analysis | TRM | C0002-4-f |

VRM: viscous remanent magnetization; IRM: isothermal remanent magnetization; TRM: thermoremanent magnetization.

**Table S2.** Summary of experimental samples in Nakamura et al. (2022)

| Series | Sample | Measurement sequence of this study |
|---|---|---|
| Initial analysis | A0026[*] | THD and stepwise AFD of NRM and IRM |
| | C0002-4-f[*] | THD and stepwise AFD of NRM and IRM |

[*]Magnetic measurements were conducted in the initial comprehensive analysis (Nakamura et al., 2022).

**Table S3.** Summary of paleomagnetic and paleointensity re-analysis for Ryugu samples in Nakamura et al. (2022)

| Sample | Linear segment (mT) | MAD (Degree) | AF segment (mT) | N | Slope | R | Palaeointensity (μT) |
|---|---|---|---|---|---|---|---|
| A0026 | 10–32.5 | 5.9 | 10–32.5 | 12 | 0.0236 ± 0.0016 | 0.977 | 78.3 ± 5.4 |
| C0002-4-f | 0[*]–22 | 2.1 | 14–22 | 5 | 0.0191 ± 0.0030 | 0.965 | 63.2 ± 9.9 |

[*]Natural remanent magnetization after 110°C thermal demagnetization.